\documentclass[10pt,twocolumn,journal]{IEEEtran}

\usepackage{amsmath}
\usepackage{amssymb}
\usepackage{amsfonts}
\usepackage{mathrsfs}
\usepackage{mathtools}
\usepackage{graphicx}
\usepackage[T1]{fontenc}
\usepackage{bm}
\usepackage[export]{adjustbox}
\usepackage[protrusion=true, expansion=true]{microtype}
\usepackage{soul}
\usepackage{cite}
\usepackage[
    colorlinks=true, 
    linkcolor=blue,  
    urlcolor=black,  
    bookmarks=false, 
] {hyperref}

\graphicspath{{./fig/}}
\title{Efficient Tracking of Sparse Signals via an Earth Mover's Distance Dynamics Regularizer}

\author
{Nicholas~P.~Bertrand$^*$\thanks{$^*$Equal contributions.},
Adam~S.~Charles$^*$,~\IEEEmembership{Member,~IEEE,}
John~Lee$^*$,
\thanks{
    N. P. Bertrand, J. Lee, P. B. Dunn, and C. J. Rozell are with the School of Electrical and Computer Engineering, Georgia Institute of Technology, Atlanta, GA 30332-0250 USA (email: nbertrand@gatech.edu; john.lee@gatech.edu; paveldunn@gatech.edu; crozell@gatech.edu). A. S. Charles is with the Princeton Neuroscience Institute, Princeton University, Princeton, NJ 08540 USA (email: adamsc@princeton.edu).
}
Pavel~B.~Dunn,
Christopher~J.~Rozell~\IEEEmembership{Senior Member,~IEEE}
\thanks{
This work was partially supported by NSF grant CCF-1409422, James S. McDonnell Foundation grant number 220020399, NIH NRSA Training Grant in Quantitative Neuroscience number T32MH065214 and the DSO National Laboratories of Singapore.
}
}
\date{}

\begin{document}
\maketitle

\begin{abstract}
Tracking algorithms such as the Kalman filter aim to improve inference performance by leveraging the temporal dynamics in streaming observations.
However, the tracking regularizers are often based on the $\ell_p$-norm which cannot account for important geometrical relationships between neighboring signal elements.
We propose a practical approach to using the earth mover's distance (EMD) via the earth mover's distance dynamic filtering (EMD-DF) algorithm for causally tracking time-varying sparse signals when there is a natural geometry to the coefficient space that should be respected (e.g., meaningful ordering).
Specifically, this paper presents a new Beckmann formulation that dramatically reduces computational complexity, as well as an evaluation of the performance and complexity of the proposed approach in imaging and frequency tracking applications with real and simulated neurophysiology data.
\end{abstract}

\section{Introduction}

Tracking algorithms aim to improve the performance of statistical inference procedures for time series by incorporating information from a dynamics model that describes how the signal evolves.
We consider the linear observation model
\begin{equation}
\bm{y}_n = \bm{A}_n \bm{x}_n + \sigma \bm{\epsilon}_n,
\end{equation}
where for each time step $n$, $\bm{x}_n$ is the underlying signal, $\bm{A}_n$ is a linear observation operator, $\sigma \bm{\epsilon}_n$ is Gaussian measurement noise with variance $\sigma^2$, and $\bm{y}_n$ is the resulting measurement vector.
The signal evolves according to a dynamics function $g$:
\begin{equation}
\bm{x}_{n+1} = g_n(\bm{x}_n) + \bm{\eta}_n,
\end{equation}
where $\bm{\eta}_n$ is a noise vector called the \emph{innovations} that accounts for inaccurate modeling of the dynamics.
When $g$ is linear and the signal, observation noise and innovations are Gaussian, the classical Kalman filter provides an efficient way to compute the optimal (i.e., minimum expected $\ell_2$ error) estimate taking into account all measurements up to the current time  \cite{Kalman1960NewApproach}.
The estimate produced by Kalman filtering may be expressed as
\begin{equation}
\widehat{\bm{x}}_n =
\operatorname*{argmin}_{\bm{x}}
\left\lVert\bm{y}_n - \bm{A}_n \bm{x}\right\rVert_{\bm{B}}^2
+ \left\lVert\bm{x} - \bm{G}_{n-1}\widehat{\bm{x}}_{n-1}\right\rVert^2_{\bm{C}},
\end{equation}
where $\bm{G}_n$ is the matrix representing the dynamics operator $g_n$, $\bm{\widehat{x}}_{n-1}$ is the previous signal estimate, and $\left\lVert\cdot\right\rVert_{\bm{B}}^2$ and $\left\lVert\cdot\right\rVert_{\bm{C}}^2$ denote Mahalanobis norms weighted appropriately by the covariance matrices of the noise, innovations, and previous signal estimate.
The Kalman filter and its extensions \cite{Haykin2001KalmanNN} have been used exhaustively in a multitude of scientific and engineering applications.

In addition to these classic models, non-Gaussian sparsity models have become increasingly popular due to their state-of-the-art performance in a variety of problems (e.g., in image processing \cite{Elad2010RedundantRepresentations} and compressive sensing \cite{Baraniuk2007CS}).
Sparse inference problems with static data vectors are well-studied, resulting in many algorithmic advances and performance guarantees~\cite{Baraniuk2007CS,Tropp2007OMP}.
In the spirit of the Kalman filter, sparse tracking algorithms have shown utility for dynamic filtering with time-varying sparse signals \cite{Asif2010DynamicUpdating, Asif2011EstimationDynamicUpdating, Ziniel2010TrackingBeliefPropagation, Vaswani2010ModifiedCS,  Charles2011SparsityPenalties, Hall2013DynamicalModels, Charles2016DynamicFiltering, balavoine2015discrete}.
However, in many applications with discretized domains, commonly used pointwise dynamics regularizers (e.g., the $\ell_p$-norm) disproportionately penalize predictions with slight mismatch in the signal support because they do not incorporate knowledge of meaningful geometry (when it exists) into the penalty.
Consider, for example, an imaging scenario where we wish to track a single-pixel target moving through a scene.
An $\ell_p$-norm based regularizer assigns equal penalties to any prediction in which the target is not precisely in the correct support location regardless of the distance between the erroneous pixel and the true position.
Similarly, when tracking time varying frequencies, the ordering of the frequencies in the discrete Fourier transform (DFT) matrix results in a geometric relationship among the DFT coefficients which is not effectively utilized with $\ell_p$-norm regularizers.
Although one may consider tracking in coordinate space instead of pixel space, this approach scales poorly in the number of targets.

In this work, we propose a practical approach to using the earth mover's distance (EMD) via the earth mover's distance dynamic filtering (EMD-DF) algorithm for causally tracking time-varying sparse signals when there is a natural geometry to the coefficient space that should be respected (e.g., meaningful ordering).
Specifically, this paper presents a new Beckmann formulation that dramatically reduces computational complexity, as well as an evaluation of the performance and complexity of the proposed approach in imaging and frequency tracking applications with real and simulated neurophysiology data.
The novel algorithmic formulation, performance characterization, and evaluation on real data introduced in this work represent crucial developments in the practicality of our approach beyond the preliminary explorations presented in~\cite{Charles2017emd, Bertrand2018emd}.

\section{Background}

\subsection{Sparse Dynamic Filtering}
A vector $\bm{x} \in \mathbb{C}^N$ is said to be sparse if only a few of its elements are non-zero.
Suppose $\bm{y}$ contains noisy observations of $\bm{x}$ through a linear measurement operator $\bm{A} \in \mathbb{C}^{M \times N}$.
For example, results in the compressed sensing literature show that under certain conditions on $\bm{A}$, $\bm{x}$ may be recovered from $\bm{y}$ even when $M \ll N$.
Of the many sparse inverse algorithms that exist (e.g., \cite{Tropp2007OMP, Candes2005L1Magic}), one popular optimization-based approach is Basis-Pursuit Denoising (BPDN):
\begin{equation}
\widehat{\bm{x}} = \operatorname*{argmin}_{\bm{x}} \enspace
\frac{1}{2}\left\lVert\bm{y} - \bm{A}\bm{x}\right\rVert_2^2 +
\lambda \left\lVert\bm{x}\right\rVert_1.
\end{equation}

Recent work has also extended these ideas for static sparse recovery to tracking algorithms for sparse time-varying signals.
Early work in this area included batch (i.e., non-causal) approaches \cite{Sankaranarayanan2010CompressiveDynamicScenes, Asif2013MotionAdaptiveMRI} and modifications to the causal Kalman filter \cite{Carmi2010SparseKalman}.
More recent causal approaches include Basis Pursuit Denoising Dynamic Filtering (BPDN-DF) which provides theoretical convergence guarantees, and Reweighted-$\ell_1$ Dynamic Filtering (RWL1-DF) which was found to be more robust to model mismatch \cite{Charles2016DynamicFiltering}.

BPDN-DF modifies standard BPDN with the addition of a tracking regularizer:
\begin{equation}
\widehat{\bm{x}}_n = \operatorname*{argmin}_{\bm{x}} \enspace
\frac{1}{2}\left\lVert\bm{y}_n - \bm{A}\bm{x}\right\rVert_2^2
+
\lambda \left\lVert\bm{x}\right\rVert_1
+
\gamma \left\lVert\bm{x} - \widetilde{\bm{x}}_n\right\rVert_2^2,
\end{equation}
where $\widetilde{\bm{x}}_n = g(\widehat{\bm{x}}_{n-1})$ is the prediction produced using the dynamics function $g$.
This additional term encourages solutions which adhere to the dynamics model.
Similarly, RWL1-DF modifies RWL1 by injecting dynamics into the recovery process via second order statistics.
While both BPDN-DF and RWL1-DF can improve performance in the recovery of time varying signals, each of these algorithms injects dynamics information in a point-wise fashion and thus fail to capture the geometric relationship between neighboring signal elements.

\subsection{Earth Mover's Distance}
\label{sec:EMD}
The earth mover's distance (EMD) is a metric which grew out of the optimal transport (OT) literature initiated by Monge \cite{Monge1781Memoire}.
The EMD has recently been increasingly used in a variety of applications such as image and histogram comparison \cite{Rubner2000EMDImageRetrieval, Ling2007EMDHistogram}, as well as for sparse inverse problems \cite{Gupta2010SparseEMD, Schmidt2013ConstrainedEMD, Mo2013CompressiveParamEstimation}.
Intuitively, if we visualize the first signal as being composed of piles of dirt and the second as holes, the EMD computes the minimum amount of \emph{work} needed to fill the holes with dirt%
\footnote{We refer the reader to \cite{Bertrand2018emd} for mathematical details of the traditional EMD formulation.}.
A key property to note is that the EMD is inherently aware of the geometric relationship between signal elements via a user-defined distance matrix $(R_{ij})$ which describes the cost to transport mass along the signal support.
This is in stark contrast to $\ell_p$ metrics, and is the primary motivation for its use as a tracking regularizer.

The traditional EMD formulation involves solving for $\mathcal{O}\left(N^2\right)$ flow variables, which has the potential to be computationally prohibitive for large problems.
For applications where the cost matrix $\bm{R}$ represents Euclidean distances (e.g., video with uniformly gridded pixels), geometric structure can be exploited to also reduce the optimization variable complexity in exact EMD solutions from $\mathcal{O}\left(N^2\right)$ to $\mathcal{O}\left(N\right)$ via the \emph{Beckmann problem} \cite{beckmann1952continuous, li2018parallel}.
This formulation poses the EMD as a minimum flux problem of a fluid flowing between a source and a sink (i.e., the input arguments of the EMD problem):
\begin{alignat}{1} \label{eqn:beckmann_emd}
d_\text{emd}\left(\bm{x},\bm{y}\right) = \min_{\bm{M}} &\enspace \| \bm{M} \|_{2,1} \nonumber \\
\quad \text{subject to} &\enspace
\operatorname{div}(\bm{M}) + \bm{y} - \bm{x} = \bm{0} ,
\end{alignat}
where the divergence operator is defined as
\begin{alignat}{1}
  \operatorname{div}(\bm{M})[i,j] = &\enspace (M_x[i,j] - M_x[i-1,j]) \nonumber \\
   + &\enspace(M_y[i,j] - M_y[i,j-1]),
\end{alignat}
the rows of $\bm{M}$ contain points in a $D$-dimensional vector field, $\| \bm{M} \|_{2,1} := \sum_{i=1}^N \| \bm{m}_i \|_2$ denotes the sum of their Euclidean norms and zero-flux boundary conditions are enforced (i.e., $M[i,j]=0$ whenever $i$ or $j$ falls outside the support).


\section{Earth Mover's Distance Dynamic Filtering}
\label{sec:EMD-DF}
In earth mover's distance dynamic filtering (EMD-DF), the causal estimate of the signal at time $n$ is given by:
\begin{equation}
\label{eqn:bpdndf_to_emddf}
\widehat{\bm{x}}_n = \operatorname*{argmin}_{\bm{x}} \enspace
\frac{1}{2}\left\lVert\bm{y}_n - \bm{A}\bm{x}\right\rVert_2^2
+
\lambda \left\lVert\bm{x}\right\rVert_1
+
\gamma d_\text{emd}\left(\bm{x},\widetilde{\bm{x}}_n\right).
\end{equation}
EMD-DF has a similar structural form as BPDN-DF at first glance~\cite{Charles2016DynamicFiltering,Charles2017emd, Bertrand2018emd}, but the use of an EMD penalty instead of an $\ell_2$ dynamics regularizer is non-trivial because the evaluation of the EMD itself requires the solution of an optimization program.
Using the traditional EMD formulation for general cost distances, the EMD-DF optimization program \eqref{eqn:bpdndf_to_emddf} involves solving $N$ signal variables and an additional $N^2$ flow variables.
Thus, EMD regularization incurs a potentially prohibitive increase in computational complexity compared to algorithms such as BPDN or RWL1.
However, in the common case when the transport costs are Euclidean (i.e., $R_{ij}=\left\lVert z_i-z_j\right\rVert_2^2$ where $z_i$ and $z_j$ are the support locations of signal elements $x_i$ and $x_j$), we can exploit Beckmann's formulation of the EMD \eqref{eqn:beckmann_emd} to reduce the number of EMD variables from $\mathcal{O}\left(N^2\right)$ to $\mathcal{O}\left(N\right)$.
This critical reduction in computational complexity enables previously intractable applications.
Preliminary explorations \cite{Charles2017emd, Bertrand2018emd} have previously addressed traditional EMD limitations to account for signals that are signed or complex valued.

The Beckmann EMD requires that the signals have unit mass (i.e., $\left\lVert\bm{x}\right\rVert_1 = \left\lVert\bm{y}\right\rVert_1 = 1$), meaning that we cannot simply apply the pre-existing method.
In the following, we outline how a reformulation of the Beckmann problem for unequal total masses \cite{Ryu2017Unbalanced} may be incorporated into the EMD-DF program.
To allow input arguments with unequal total mass, we introduce slack variables $\bm{w},\bm{v}$ to bound the flux from the original source $\bm{x}$ and sink $\bm{y}$.
The modified EMD program is then:
\begin{alignat}{1} \label{eqn:beckmann_emd_unbalanced}
d_\text{emd}\left(\bm{x},\bm{y}\right) = \min_{\bm{M},\bm{w},\bm{v}} &\enspace \| \bm{M} \|_{2,1} \nonumber \\
\quad \text{subject to} &\enspace
\operatorname{div}(\bm{M}) + \bm{v} - \bm{w} = \bm{0} , \nonumber \\
&\enspace
\bm{0} \leq \bm{w} \leq \bm{x}, \bm{0} \leq \bm{v} \leq \bm{y} , \nonumber \\
&\enspace
\left\lVert\bm{w}\right\rVert_1 = \left\lVert\bm{v}\right\rVert_1 = \min(\left\lVert\bm{x}\right\rVert_1,\left\lVert\bm{y}\right\rVert_1) ,
\end{alignat}
where $\bm{w},\bm{v}$ are nonnegative vectors with the same dimensions as $\bm{x},\bm{y}$.
This optimization searches for the minimal vector field configuration that describes, via the first constraint, its flux to be traveling between a source $\bm{w}$ and a sink $\bm{v}$.
The second constraint describes the source and sink as nonnegative slack variables that are bounded above by their proxies $\bm{x}$ and $\bm{y}$ respectively; this constraint is analogous to the mass preservation constraints in the traditional EMD formulation.
The last constraint states that the induced flux must be bounded by the total mass of the smaller operand signal.
This formulation has $N(D+2)$ variables, where $D$ is the dimensions of the vector field (e.g., $D=2$ for images).
Applying this EMD formulation and employing the same strategy as \cite{Bertrand2018emd} to replace the $\min$ term with a slack variable $u$, \eqref{eqn:bpdndf_to_emddf} becomes
    \begin{alignat}{1} \label{eqn:beckmann_slack_eqn}
    \widehat{\bm{x}}_n = \operatorname*{argmin}_{\bm{x},\bm{M},u,\bm{v},\widetilde{\bm{v}}} &\enspace
    \frac{1}{2}\left\lVert\bm{y}_n - \bm{A}\bm{x}\right\rVert_2^2
    + \lambda \left\lVert\bm{x}\right\rVert_1 \nonumber \\
    &\enspace
    + \gamma \| \bm{M} \|_{2,1}
    - \mu u \nonumber \\
    \text{subject to} &\enspace
    \operatorname{div}(\bm{M}) + \widetilde{\bm{v}} - \bm{v} = \bm{0}, \nonumber \\
    &\enspace
    \bm{0} \leq \bm{v} \leq \bm{x}, \enspace
    \bm{0} \leq \widetilde{\bm{v}} \leq \widetilde{\bm{x}}, \nonumber \\
    &\enspace
    \left\lVert\bm{v}\right\rVert_1 = \left\lVert\widetilde{\bm{v}}\right\rVert_1 = u, \nonumber \\
    &\enspace
    u \leq \left\lVert\bm{x}\right\rVert_1, \enspace
    u \leq \left\lVert\widetilde{\bm{x}}\right\rVert_1.
    \end{alignat}
The complex variant of EMD-DF in \cite{Bertrand2018emd} can also be trivially converted to adopt this formulation, though it is not shown here for the sake of brevity.
As we mentioned in Section~\ref{sec:EMD}, \eqref{eqn:beckmann_slack_eqn} enjoys a reduction in variable complexity from $\mathcal{O}\left(N^2\right)$ to $\mathcal{O}\left(N\right)$ while preserving the benefits of partial OT (in contrast to the traditional balanced OT Beckmann formulation) and avoiding the approximation error associated with methods such as Sinkhorn iterations.

Finally, we note that other recent works \cite{Karlsson2017GeneralizedSinkhornIterations, Janati2019GroupLevelMEG, Janati2018WassersteinRegularizationSparse} incorporate optimal transport regularizers in inverse problems using the Sinkhorn algorithm \cite{Sinkhorn1964RelationshipArbitraryPositive,Cuturi2013SinkhornDistancesLightspeed}.
However, the work presented here is distinct in two subtle but important ways.
First, our proposed partial Beckmann formulation provides an alternative numerical approach that offers attractive linear variable complexity (similar to Sinkhorn methods) but without sacrificing accuracy.
Sinkhorn approaches use entropic regularization to trade off accuracy vs. speed, limiting their utility for finding sparse solutions over time due to mass diffusion across neighboring support. In contrast, the Beckmann formulation reflects the true optimal transport distance (subject only to negligible discretization errors).
Second, our partial transport formulation results in a linearly constrained quadratic program that is easily implementable with off-the-shelf solvers (e.g., CVX, Gurobi, Mosek).
In contrast, an unbalanced Sinkhorn approach would necessitate a custom solver (e.g., alternating optimization \cite{janati2019group, janati2018wasserstein}) that would be non-trivial to implement.

\section{Results}
We demonstrate the utility and performance of EMD-DF through a series of simulations on synthetic and real data.
First, we study the problem of tracking time varying frequencies in a 1-D time series of real neurophysiology data.
Next, we consider the problem of tracking a wavefront in synthetically generated data motivated by the phenomenon of traveling waves which appear in electrophysiology data.
Finally, we demonstrate the significant numerical speed up of EMD-DF due to Beckmann's formulation.
Throughout these simulations, we use the CVX software package \cite{Grant2014CVX} which employs interior point methods to carry out the EMD-DF optimization.
Hyperparameters are tuned manually for  real data, while direct search \cite{Kolda2003OptimizationDirectSearch} is used for synthetic data where ground truth is known.

\subsection{Tracking Neural Oscillations}
In this section, we apply EMD-DF to the problem of spectrum estimation in neurophysiology  recordings.
Oscillatory behavior is prominent in a variety of neural recording settings and there is great interest in the neuroscience community to understand the functional role of these oscillations \cite{Cornelissen2015AgedependentElectroencephalogramEEG,Aru2015UntanglingCrossfrequencyCoupling}.

In many studies, the tools used for spectral analysis of neural recordings are based on the classical short-time Fourier transform (STFT).
The time and frequency resolution of such techniques is thus limited by the uncertainty principle which prevents simultaneously achieving high frequency and time resolution.
Here, we study how higher time-frequency (TF) resolution may be obtained by imposing a sparsity model on the data and using EMD-DF for recovery in an overcomplete DFT dictionary.

\begin{figure}[t]
  \centering
  \includegraphics{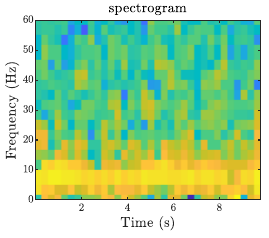}
  \includegraphics{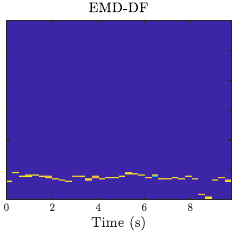}
\caption{
    Time-frequency plots for a single channel of tetrode data recorded from the rat hippocampus.
    Data is sampled at 250 Hz and an analysis window length of 72 samples is used for both plots.
    The spectrogram (left), which is produced using the traditional STFT with a hamming window, yields lower frequency resolution and severe leakage in the lower frequencies.
    The TF plot on the right is produced by EMD-DF with an 5 times overcomplete DFT matrix, resulting in high enough frequency resolution to smoothly track subtle changes in frequency.
}
\label{fig:kemere}
\end{figure}

EMD-DF may be used as a causal alternative to well-established spectral sharpening methods \cite{Gardner2006SparseTF, Thomson2000Multitaper, Auger1995ImprovingReadabilityTF, Fulop2006ReassignedSpectrogram} which also aim to improve the resolution of time-frequency representations, but are restricted to operate on batch data \cite{Bertrand2018emd}.
Causal algorithms are crucial in online applications such as closed-loop control.
Here we employ EMD-DF to estimate the spectrum in a segment of real electrophysiology data recorded from a tetrode in rat hippocampus \cite{Kemere2013Hippocampal}.
\footnote{ The authors would like to thank C.~Kemere for the tetrode recording data.}
We take the measurement matrix $\bm{A}$ to be a 5 times overcomplete DFT matrix and track the top two frequencies by setting the remaining frequencies in the prediction to zero via the dynamics function $g$.
Figure~\ref{fig:kemere} shows TF plots produced by the spectrogram and EMD-DF.
Because EMD-DF utilizes the overcomplete DFT matrix for recovery, it produces a TF plot with vastly improved frequency resolution.
Additionally, the spectrogram suffers from severe leakage in the 5--10Hz frequency band, an artifact which is not present in the sparse TF representation.
Finally, the improved resolution of the sparse TF plot reveals more subtle oscillatory dynamics that cannot be observed in the spectrogram.

\subsection{Tracking Traveling Waves}
\begin{figure}[t]
  \centering
  \includegraphics{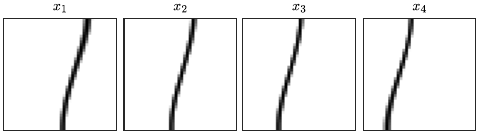}\\
  \includegraphics{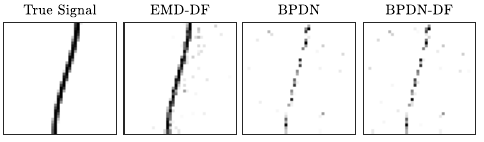}\\~\\
  \includegraphics{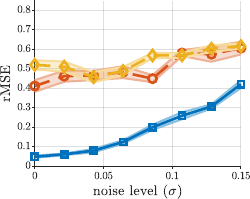}
  \includegraphics{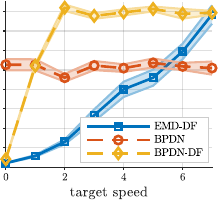}
\caption{
Single step recovery of wavefronts from a $40\times 40$ Kuramoto oscillator array via linear Gaussian measurements.
(Top) A linear gradient in the oscillators' intrinsic frequencies results in traveling waves across the array.
(Middle) Examples of wavefronts recovered using several methods.
BPDN produces poor recovery performance due to undersampled ($M/N=0.15$) and noisy ($\sigma=0.08$) measurements.
Dynamics regularization in BPDN-DF is of little help due to its inability to effectively utilize the prediction which has little support overlap with the ground truth signal.
EMD regularization is robust to the mismatch in pixel location between the ground truth and prediction and thus enables successful recovery of the wavefront.
(Bottom row) Recovery performance averaged over 10 trials.
Error bars indicate $\alpha=0.01$ confidence intervals.
(Bottom-left) EMD-DF produces superior performance for various values of the noise standard deviation $\sigma$.
(Bottom-right) The $\ell_2$-norm dynamics regularization in BPDN-DF actually degrades performance compared to BPDN for moving targets.
In contrast, EMD-DF is substantially more robust to support location mismatch caused by target movement.
}
\label{fig:kuramoto}
\end{figure}

Traveling waves are another form of neural oscillation pattern of interest in the neuroscience community.
For example, wave propagation has been shown to correlate to events and performance in tasks involving neurosurgical patients \cite{Zhang2018ThetaAlphaOscillations}.
We generate synthetic traveling wave data using the phase-coupled Kuramoto oscillator model which has been used to study the properties of traveling waves in neuronal activity \cite{Ermentrout2001TravelingElectricalWaves,Kuramoto1981RhythmsTurbulencePopulations}.
The Kuramoto model describes the instantaneous phase of each node in an array of linked oscillators via a system of differential equations.
Motivated by the work in \cite{Ermentrout2001TravelingElectricalWaves}, we simulate a $40 \times 40$ oscillator array where the instantaneous phase of oscillator $(i,j)$ with intrinsic frequency $\omega_{ij}$ depends on its four nearest neighbors via the equation $\theta_{ij}' = \omega_{ij} + 300\sum_{k=1}^4 \sin(\theta_k)$.
We choose $\omega_{ij} = 2 + 0.103i + 0.359j$ to produce a linear frequency gradient which has been observed to yield traveling wave solutions.
After solving for the $\theta_{ij}$, we threshold the oscillator voltage $\sin(\theta_{ij}(t))$ to extract the wavefront $x_{ij}(t)$.
The goal of these simulations is to recover the wavefront from noisy linear measurements $\bm{y}(t) = \bm{\Phi}\bm{x}(t) + \bm{\epsilon}$.
We measure upper bound tracking performance by providing the ground truth previous frame as the prediction.
Figure~\ref{fig:kuramoto} shows how EMD-DF enables nearly perfect recovery compared to BPDN and BPDN-DF which are unable to effectively use the prediction.

\subsection{Computational Scalability}
\begin{figure}[t]
  \centering
  \includegraphics{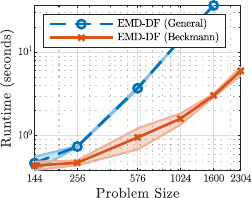}
  \includegraphics{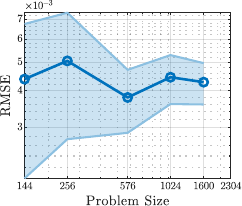}
\caption{
  Evaluation of computational speed up.
  We compare the runtime and the difference in solutions for two formulations of EMD-DF: EMD-DF (General) which adopts generic distance costs, and EMD-DF (Beckmann) which assumes Euclidean distance costs.
  The left plot demonstrates that EMD-DF (Beckmann) significantly outperforms EMD-DF (General) in runtime, and in the right plot, the differences in solutions are shown to be negligible.
}
\label{fig:computational}
\end{figure}

Finally, we evaluate the runtime of Beckmann EMD-DF compared to previously developed versions which use the traditional EMD formulation with general ground costs $R_{ij}$.
We conduct a stylized target tracking simulation in which a sparse collection of targets move to adjacent support locations with equal probability.
We take random Gaussian measurements and scale the problem between state sizes of $12 \times 12$ $(N=144)$ and $48 \times 48$ $(N=2304)$.
For each state size, the sparsity level is fixed at 5\% and 10 trials are run on a personal computer with a 3.5 Ghz Intel Core i7 processor.
The Beckmann formulation yields solutions significantly faster than the general EMD formulation, especially for large problem sizes (Figure~\ref{fig:computational}).
We also note that the discrepancy between solutions obtained using each method (measured in root mean squared error: $\left\lVert\bm{x}-\bm{y}\right\rVert_2^2/\sqrt{N}$) is negligible --- on the order of $10^{-3}$ throughout the entire range of problem sizes.

\section{Summary}
EMD-DF can be a more effective dynamics regularizer for sparse dynamic filtering compared to traditional point-wise methods (e.g., the $\ell_p$-norm), but potentially suffers from computational complexity that limits problem sizes.
The results presented here demonstrate that EMD-DF is a practical sparse tracking method when used with a Beckmann formulation that scales linearly in the number of optimization variables instead of quadratically, enabling substantial performance gains for large-scale problems.
Finally, we demonstrate the utility of EMD-DF by tracking wavefronts in a Kuramoto oscillator network and by tracking sparse frequencies in real electrophysiology data.

\bibliographystyle{IEEEtran}
\bibliography{refs}

\begin{thebibliography}{10}
\providecommand{\url}[1]{#1}
\csname url@samestyle\endcsname
\providecommand{\newblock}{\relax}
\providecommand{\bibinfo}[2]{#2}
\providecommand{\BIBentrySTDinterwordspacing}{\spaceskip=0pt\relax}
\providecommand{\BIBentryALTinterwordstretchfactor}{4}
\providecommand{\BIBentryALTinterwordspacing}{\spaceskip=\fontdimen2\font plus
\BIBentryALTinterwordstretchfactor\fontdimen3\font minus
  \fontdimen4\font\relax}
\providecommand{\BIBforeignlanguage}[2]{{%
\expandafter\ifx\csname l@#1\endcsname\relax
\typeout{** WARNING: IEEEtran.bst: No hyphenation pattern has been}%
\typeout{** loaded for the language `#1'. Using the pattern for}%
\typeout{** the default language instead.}%
\else
\language=\csname l@#1\endcsname
\fi
#2}}
\providecommand{\BIBdecl}{\relax}
\BIBdecl

\bibitem{Kalman1960NewApproach}
R.~E. Kalman, ``A new approach to linear filtering and prediction problems,''
  \emph{Journal of Basic Engineering}, vol.~82, no.~1, pp. 35--45, Mar. 1960.

\bibitem{Haykin2001KalmanNN}
S.~S. Haykin, \emph{\BIBforeignlanguage{English}{Kalman Filtering and Neural
  Networks}}.\hskip 1em plus 0.5em minus 0.4em\relax New York: {Wiley}, 2001,
  oCLC: 52366672.

\bibitem{Elad2010RedundantRepresentations}
M.~Elad, M.~A.~T. Figueiredo, and Y.~Ma, ``On the role of sparse and redundant
  representations in image processing,'' \emph{Proceedings of the IEEE},
  vol.~98, no.~6, pp. 972--982, Jun. 2010.

\bibitem{Baraniuk2007CS}
R.~G. Baraniuk, ``Compressive sensing [lecture notes],'' \emph{IEEE Signal
  Processing Magazine}, vol.~24, no.~4, pp. 118--121, Jul. 2007.

\bibitem{Tropp2007OMP}
J.~A. Tropp and A.~C. Gilbert, ``Signal recovery from random measurements via
  orthogonal matching pursuit,'' \emph{IEEE Transactions on Information
  Theory}, vol.~53, no.~12, pp. 4655--4666, Dec. 2007.

\bibitem{Asif2010DynamicUpdating}
M.~S. Asif and J.~Romberg, ``Dynamic updating for $\ell_1$ minimization,''
  \emph{IEEE Journal of Selected Topics in Signal Processing}, vol.~4, no.~2,
  pp. 421--434, Apr. 2010.

\bibitem{Asif2011EstimationDynamicUpdating}
M.~S. Asif, A.~Charles, J.~Romberg, and C.~Rozell, ``Estimation and dynamic
  updating of time-varying signals with sparse variations,'' in \emph{2011
  {{IEEE International Conference}} on {{Acoustics}}, {{Speech}} and {{Signal
  Processing}} ({{ICASSP}})}, May 2011, pp. 3908--3911.

\bibitem{Ziniel2010TrackingBeliefPropagation}
J.~Ziniel, L.~C. Potter, and P.~Schniter, ``Tracking and smoothing of
  time-varying sparse signals via approximate belief propagation,'' in
  \emph{2010 {{Conference Record}} of the {{Forty Fourth Asilomar Conference}}
  on {{Signals}}, {{Systems}} and {{Computers}}}, Nov. 2010, pp. 808--812.

\bibitem{Vaswani2010ModifiedCS}
N.~Vaswani and W.~Lu, ``Modified-{{CS}}: Modifying compressive sensing for
  problems with partially known support,'' \emph{IEEE Transactions on Signal
  Processing}, vol.~58, no.~9, pp. 4595--4607, Sep. 2010.

\bibitem{Charles2011SparsityPenalties}
A.~Charles, M.~S. Asif, J.~Romberg, and C.~Rozell, ``Sparsity penalties in
  dynamical system estimation,'' in \emph{2011 45th {{Annual Conference}} on
  {{Information Sciences}} and {{Systems}}}, Mar. 2011, pp. 1--6.

\bibitem{Hall2013DynamicalModels}
E.~C. Hall and R.~M. Willett, ``Dynamical models and tracking regret in online
  convex programming,'' in \emph{Proceedings of the 30th {{International
  Conference}} on {{Machine Learning}}}, S.~Dasgupta and D.~McAllester,
  Eds.\hskip 1em plus 0.5em minus 0.4em\relax {PMLR}, Feb. 2013, pp. 579--587.

\bibitem{Charles2016DynamicFiltering}
A.~S. Charles, A.~Balavoine, and C.~J. Rozell, ``Dynamic filtering of
  time-varying sparse signals via $\ell _1$ minimization,'' \emph{IEEE
  Transactions on Signal Processing}, vol.~64, no.~21, pp. 5644--5656, Nov.
  2016.

\bibitem{balavoine2015discrete}
A.~Balavoine, C.~J. Rozell, and J.~Romberg, ``Discrete and continuous-time
  soft-thresholding for dynamic signal recovery,'' \emph{IEEE Transactions on
  Signal Processing}, vol.~63, no.~12, pp. 3165--3176, 2015.

\bibitem{Charles2017emd}
A.~S. Charles, N.~P. Bertrand, J.~Lee, and C.~J. Rozell, ``Earth-mover's
  distance as a tracking regularizer,'' in \emph{2017 IEEE 7th International
  Workshop on Computational Advances in Multi-Sensor Adaptive Processing
  (CAMSAP)}, Dec 2017, pp. 1--5.

\bibitem{Bertrand2018emd}
N.~P. Bertrand, J.~Lee, A.~S. Charles, P.~Dunn, and C.~J. Rozell, ``Sparse
  dynamic filtering via earth mover's distance regularization,'' in
  \emph{Proceedings of the IEEE International Conference on Acoustics, Speech,
  and Signal Processing (ICASSP), Calgary, Alberta, Canada}, Apr 2018.

\bibitem{Candes2005L1Magic}
E.~Cand\`{e}s and J.~Romberg, ``$\ell_1$-magic: recovery of sparse signals via
  convex programming,'' California Institute of Technology, Tech. Rep., 2005.

\bibitem{Sankaranarayanan2010CompressiveDynamicScenes}
A.~C. Sankaranarayanan, P.~K. Turaga, R.~G. Baraniuk, and R.~Chellappa,
  ``\BIBforeignlanguage{en}{Compressive acquisition of dynamic scenes},'' in
  \emph{\BIBforeignlanguage{en}{Computer {{Vision}} \textendash{} {{ECCV}}
  2010}}, ser. Lecture Notes in Computer Science.\hskip 1em plus 0.5em minus
  0.4em\relax {Springer, Berlin, Heidelberg}, Sep. 2010, pp. 129--142.

\bibitem{Asif2013MotionAdaptiveMRI}
M.~S. Asif, L.~Hamilton, M.~Brummer, and J.~Romberg,
  ``\BIBforeignlanguage{en}{Motion-adaptive spatio-temporal regularization for
  accelerated dynamic {{MRI}}},'' \emph{\BIBforeignlanguage{en}{Magnetic
  Resonance in Medicine}}, vol.~70, no.~3, pp. 800--812, Sep. 2013.

\bibitem{Carmi2010SparseKalman}
A.~Carmi, P.~Gurfil, and D.~Kanevsky, ``Methods for sparse signal recovery
  using {{Kalman}} filtering with embedded pseudo-measurement norms and
  quasi-norms,'' \emph{IEEE Transactions on Signal Processing}, vol.~58, no.~4,
  pp. 2405--2409, Apr. 2010.

\bibitem{Monge1781Memoire}
G.~Monge, \emph{\BIBforeignlanguage{French}{{M{\'e}moire sur la th{\'e}orie des
  d{\'e}blais et des remblais}}}.\hskip 1em plus 0.5em minus 0.4em\relax Paris:
  {De l'Imprimerie Royale}, 1781, oCLC: 51928110.

\bibitem{Rubner2000EMDImageRetrieval}
Y.~Rubner, C.~Tomasi, and L.~J. Guibas, ``\BIBforeignlanguage{en}{The earth
  mover's distance as a metric for image retrieval},''
  \emph{\BIBforeignlanguage{en}{International Journal of Computer Vision}},
  vol.~40, no.~2, pp. 99--121, Nov. 2000.

\bibitem{Ling2007EMDHistogram}
H.~Ling and K.~Okada, ``An efficient earth mover's distance algorithm for
  robust histogram comparison,'' \emph{IEEE Transactions on Pattern Analysis
  and Machine Intelligence}, vol.~29, no.~5, pp. 840--853, May 2007.

\bibitem{Gupta2010SparseEMD}
R.~Gupta, P.~Indyk, and E.~Price, ``Sparse recovery for earth mover distance,''
  in \emph{2010 48th {{Annual Allerton Conference}} on {{Communication}},
  {{Control}}, and {{Computing}} ({{Allerton}})}, Sep. 2010, pp. 1742--1744.

\bibitem{Schmidt2013ConstrainedEMD}
L.~Schmidt, C.~Hegde, and P.~Indyk, ``The constrained earth mover distance
  model, with applications to compressive sensing,'' in \emph{10th
  {{International Conference}} on {{Sampling Theory}} and {{Applications}}
  ({{SAMPTA}})}, 2013.

\bibitem{Mo2013CompressiveParamEstimation}
D.~Mo and M.~F. Duarte, ``Compressive parameter estimation with earth mover's
  distance via {{K}}-median clustering,'' in \emph{Proc. {{SPIE}} 8858,
  {{Wavelets}} and {{Sparsity XV}}}, vol. 8858, 2013, pp. 88\,581P--8858--10.

\bibitem{beckmann1952continuous}
M.~Beckmann, ``A continuous model of transportation,'' \emph{Econometrica:
  Journal of the Econometric Society}, pp. 643--660, 1952.

\bibitem{li2018parallel}
W.~Li, E.~K. Ryu, S.~Osher, W.~Yin, and W.~Gangbo, ``A parallel method for
  earth mover's distance,'' \emph{Journal of Scientific Computing}, vol.~75,
  no.~1, pp. 182--197, 2018.

\bibitem{Ryu2017Unbalanced}
E.~K. Ryu, W.~Li, P.~Yin, and S.~Osher, ``\BIBforeignlanguage{en}{Unbalanced
  and partial {$L_1$ Monge-Kantorovich} problem: a scalable parallel
  first-order method},'' \emph{\BIBforeignlanguage{en}{Journal of Scientific
  Computing}}, pp. 1--18, Nov. 2017.

\bibitem{Karlsson2017GeneralizedSinkhornIterations}
J.~Karlsson and A.~Ringh, ``Generalized {{Sinkhorn}} iterations for
  regularizing inverse problems using optimal mass transport,'' \emph{SIAM
  Journal on Imaging Sciences}, vol.~10, no.~4, pp. 1935--1962, Jan. 2017.

\bibitem{Janati2019GroupLevelMEG}
H.~Janati, T.~Bazeille, B.~Thirion, M.~Cuturi, and A.~Gramfort,
  ``\BIBforeignlanguage{en}{Group level {{MEG}}/{{EEG}} source imaging via
  optimal transport: Minimum {{Wasserstein}} estimates},'' in
  \emph{\BIBforeignlanguage{en}{Information {{Processing}} in {{Medical
  Imaging}}}}, ser. Lecture {{Notes}} in {{Computer Science}}, A.~C.~S. Chung,
  J.~C. Gee, P.~A. Yushkevich, and S.~Bao, Eds.\hskip 1em plus 0.5em minus
  0.4em\relax {Cham}: {Springer International Publishing}, 2019, pp. 743--754.

\bibitem{Janati2018WassersteinRegularizationSparse}
H.~Janati, M.~Cuturi, and A.~Gramfort, ``Wasserstein regularization for sparse
  multi-task regression,'' \emph{arXiv:1805.07833 [cs, stat]}, May 2018.

\bibitem{Sinkhorn1964RelationshipArbitraryPositive}
R.~Sinkhorn, ``A relationship between arbitrary positive matrices and doubly
  stochastic matrices,'' \emph{The Annals of Mathematical Statistics}, vol.~35,
  no.~2, pp. 876--879, 1964.

\bibitem{Cuturi2013SinkhornDistancesLightspeed}
M.~Cuturi, ``Sinkhorn {{Distances}}: {{Lightspeed Computation}} of {{Optimal
  Transport}},'' in \emph{Advances in {{Neural Information Processing Systems}}
  26}, C.~J.~C. Burges, L.~Bottou, M.~Welling, Z.~Ghahramani, and K.~Q.
  Weinberger, Eds.\hskip 1em plus 0.5em minus 0.4em\relax {Curran Associates,
  Inc.}, 2013, pp. 2292--2300.

\bibitem{janati2019group}
H.~Janati, T.~Bazeille, B.~Thirion, M.~Cuturi, and A.~Gramfort, ``Group level
  meg/eeg source imaging via optimal transport: minimum wasserstein
  estimates,'' in \emph{International Conference on Information Processing in
  Medical Imaging}.\hskip 1em plus 0.5em minus 0.4em\relax Springer, 2019, pp.
  743--754.

\bibitem{janati2018wasserstein}
H.~Janati, M.~Cuturi, and A.~Gramfort, ``Wasserstein regularization for sparse
  multi-task regression,'' \emph{arXiv preprint arXiv:1805.07833}, 2018.

\bibitem{Grant2014CVX}
M.~Grant and S.~Boyd, ``{{CVX}}: {{MATLAB}} software for disciplined convex
  programming, version 2.1,'' Mar. 2014.

\bibitem{Kolda2003OptimizationDirectSearch}
T.~Kolda, R.~Lewis, and V.~Torczon, ``Optimization by direct search: New
  perspectives on some classical and modern methods,'' \emph{SIAM Review},
  vol.~45, no.~3, pp. 385--482, Jan. 2003.

\bibitem{Cornelissen2015AgedependentElectroencephalogramEEG}
L.~Cornelissen, S.-E. Kim, P.~L. Purdon, E.~N. Brown, and C.~B. Berde,
  ``Age-dependent electroencephalogram ({{EEG}}) patterns during sevoflurane
  general anesthesia in infants,'' \emph{eLife}, vol.~4, p. e06513, Jun. 2015.

\bibitem{Aru2015UntanglingCrossfrequencyCoupling}
J.~Aru, J.~Aru, V.~Priesemann, M.~Wibral, L.~Lana, G.~Pipa, W.~Singer, and
  R.~Vicente, ``Untangling cross-frequency coupling in neuroscience,''
  \emph{Current Opinion in Neurobiology}, vol.~31, pp. 51--61, Apr. 2015.

\bibitem{Gardner2006SparseTF}
T.~J. Gardner and M.~O. Magnasco, ``\BIBforeignlanguage{en}{Sparse
  time-frequency representations},'' \emph{\BIBforeignlanguage{en}{Proceedings
  of the National Academy of Sciences}}, vol. 103, no.~16, pp. 6094--6099, Apr.
  2006.

\bibitem{Thomson2000Multitaper}
D.~J. Thomson, ``Multitaper analysis of nonstationary and nonlinear time series
  data,'' in \emph{Nonlinear and Nonstationary Signal Processing}, W.~J.
  Fitzgerald, Ed.\hskip 1em plus 0.5em minus 0.4em\relax Cambridge ; New York:
  {Cambridge University Press}, 2000, pp. 317--394, oCLC: ocm45339400.

\bibitem{Auger1995ImprovingReadabilityTF}
F.~Auger and P.~Flandrin, ``Improving the readability of time-frequency and
  time-scale representations by the reassignment method,'' \emph{IEEE
  Transactions on Signal Processing}, vol.~43, no.~5, pp. 1068--1089, May 1995.

\bibitem{Fulop2006ReassignedSpectrogram}
S.~A. Fulop and K.~Fitz, ``Algorithms for computing the time-corrected
  instantaneous frequency (reassigned) spectrogram, with applications,''
  \emph{The Journal of the Acoustical Society of America}, vol. 119, no.~1, pp.
  360--371, Jan. 2006.

\bibitem{Kemere2013Hippocampal}
C.~Kemere, M.~F. Carr, M.~P. Karlsson, and L.~M. Frank,
  ``\BIBforeignlanguage{en}{Rapid and continuous modulation of hippocampal
  network state during exploration of new places},''
  \emph{\BIBforeignlanguage{en}{PLOS ONE}}, vol.~8, no.~9, p. e73114, Sep.
  2013.

\bibitem{Zhang2018ThetaAlphaOscillations}
H.~Zhang, A.~J. Watrous, A.~Patel, and J.~Jacobs, ``Theta and alpha
  oscillations are traveling waves in the human neocortex,'' \emph{Neuron},
  vol.~98, no.~6, pp. 1269--1281.e4, Jun. 2018.

\bibitem{Ermentrout2001TravelingElectricalWaves}
G.~B. Ermentrout and D.~Kleinfeld, ``Traveling {{Electrical Waves}} in
  {{Cortex}}: {{Insights}} from {{Phase Dynamics}} and {{Speculation}} on a
  {{Computational Role}},'' \emph{Neuron}, vol.~29, no.~1, pp. 33--44, Jan.
  2001.

\bibitem{Kuramoto1981RhythmsTurbulencePopulations}
Y.~Kuramoto, ``Rhythms and turbulence in populations of chemical oscillators,''
  \emph{Physica A: Statistical Mechanics and its Applications}, vol. 106,
  no.~1, pp. 128--143, Mar. 1981.

\end{thebibliography}

\end{document}